# Analysis of Users' Behaviour and Adoption Trends of Social Media Payment Platforms

Mahdi H. Miraz
The Chinese University of Hong Kong
Sha Tin, Hong Kong SAR
m.miraz@ieee.org

Marie Haikel-Elsabeh
Léonard de Vinci Pôle Universitaire
Paris, French
marie.haikel_elsabeh@devinci.fr

*Abstract*— **The recent proliferation of Electronic Commerce (E-commerce) has been further escalated by multifaceted emerging payment solutions such as cryptocurrencies, mobile, peer-to-peer (P2P) and social media payment platforms. While these technological advancements are gaining tremendous popularity, mostly for their ease of use, various impediments such as security and privacy concerns, societal and cultural norms etc. forbear the users' adoption trends to some extents. This article examines the current status of the social media payment platforms as well as the projection of future adoption trends. Our research underlines the motivations and obstacles to the adoption of social media platforms.**

## I. INTRODUCTION

Money, including barter i.e. exchange of commodities, has been a very important catalyst of human history since 3000 years. In today's realm of connected economy, it is well-nigh impossible living a day without using some sort of money – cash or cashless. From barter to envisaged Facebook's Libra, from consensus based Rai (Fei) stones of Yap society to today's blockchain based cryptocurrency, from bank notes to plastic cards or even contactless wearables – money has gone through varied different forms – to satisfy the users' needs of that particular time aligning with the then available technologies.

The world has entered into a cashless era where the diffusion of technology [1], more particularly mobile payment platforms, has even reached the actions of the street beggars. In some countries, such as China, hawkers and beggars are seen to use Quick Response (QR) code based payment systems for the payment of the products they sell or the alms they earn by begging. While such diffusion and adoption trends of technology seem to be great blessings, there are inevitably more than what meet the eyes. Further research, especially in terms of legal, ethical, regulatory and privacy aspects, as well as appropriate actions by the government agents, more precisely in terms of regulatory and monitory point of view, are required.

A multitude of web, desktop and mobile applications are now widely available [2] to address different human needs including making payments or transferring funds. Web applications are being extended or customised to offer services for the users of handheld devices including smartphones. Therefore, to discuss social media payment platforms, discussion of mobile payment platforms are inevitably interrelated. Since social media platforms are simultaneously available for both desktop and mobile users, to investigate the user adoption trends of social media platforms, it is also necessary to explore how mobile payment platforms are being diffused amongst various users.

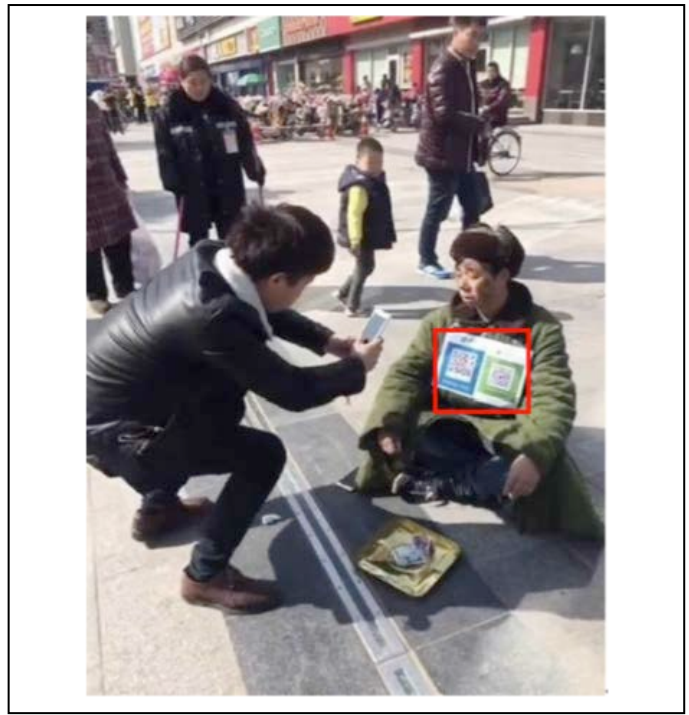

Fig. 1. Tech-savvy beggars in China[1]

## II. MOBILE PAYMENT PLATFORM TECHNOLOGIES

Amongst all other cashless payment tools, the adoption of mobile payment is developing day by day. While mobile payment is a broad term and being widely used, there is no firm definition of it. Conventionally, any payment made from or via any mobile device such as smartphones or other handheld devices can be considered as mobile payment. Instead of making the payment or transfer of money using any traditional route such as cash, cheques, cards etc., users can rather pay via various mobile payment options using their own devices generally either by using any mobile app or mobile wallet. Examples of such mobile payment platforms include: Apple Pay, PayPal Mobile, Google Wallet, Samsung Pay, WeChat Pay, Square Order, AliPay, MasterCard MasterPass, Paydiant, Intuit GoPayment, Visa Checkout, Android Pay etc. Depending on what techniques and technologies are used, there are multiple models of mobile payment. Amongst them, the following four are considered to be the major ones:

---

[1] http://www.tammyduffy.com/ARTFASHION/index.php?entry_id=2368874

### A. Short Messege Service (SMS) Payment

Payments are made via sending SMS. The recipient company has business agreement with the mobile carrier. By charging the sender at the agreed amount based on the SMS, the fund is then transferred to the recipient by the carrier. SMS charges can be prepaid for "Pay As You Go" (PAYG) or billed at the end of the regular billing period for Postpaid (Pay Monthly) or contract customers. SMS payment is very serviceable for those users who do not have a smartphone or internet connectivity.

### B. Near-field Communication (NFC) Payment

Near-field communication, more commonly known as NFC, is mainly a set of wireless communication protocols enabling two NCF enabled electronic devices to exchange data between each other. In its most commonly used scenario as a payment platform, one of the devises is a smartphone, the other is a "Point of Sale" (POS) payment terminal, to transfer funds for any goods, and services availed. Usually, the devices needs to be within a range of 4 cm to 10 cm (1.5 inch to 4 inch), depending on the Radio Frequency Identification (RFID) technology used to enable the communication. NFC payment is very popular in China, Hong Kong, Japan and many other countries. NFC is considered to be an emerging supplement of electronic ticket smart cards and even credit cards.

### C. Wireless Application Protocol (WAP) Payment

As the name implies, these models use the WAP protocol to get connected to the Internet and then make the payment, in most cases by simply clicking on the links or URLs provided. Payments can also be made using WAP protocol together with mobile wallets. WAP Payments, also known as WAP Billing, enables users to buy products or services, more particularly entertainment contents such as mobile ringtones, wallpapers and games, from businesses (websites) who has established payment partnership mobile carrier. The charges are directly added to the consumers' mobile phone bills. Thus, in most cases, the users do not require to register for the service or authenticate by providing username and password.

### D. Quick Response (QR) Code payments

QR codes of multifaceted applications including mobile payment. Use of QR code, payment or for any other applications, is becoming increasingly widespread. QR Codes for payment functions via mobile banking apps, various apps by other providers and stores, mobile wallets etc. Prior association of users card or bank details need to be associated or connected to the apps. However, p2p payment using QR codes do not require any bank account or associated card to receive funds. Payments can also be made without the need for an associated bank account or card, if there is existing balance. To withdraw the received funds, a bank account is required. If the receiver does not have a bank account, to cash the funds, they can be transferred to another third-party p2p account which is associated any bank. QR codes are also displayed in many desktop-based websites to facilitate to make payment without any data (card and other associated details) entry. Thus, the payment is not only hassle-free but also more secure as it utilises the security features of the payees' mobile carriers and apps, rather than those of the merchants' websites.

### E. Mobile Wallets

Payment information in a Mobile Wallets, also known as Digital Wallets, is commonly stored in a mobile device, more commonly in the wallet (an app) installed on the device. Apart from bank account or credit card details, various types of tickets, royalty cards, boarding passes etc. can also be associated and stored. Mobile Wallets utilise other payment protocols such as NFC and QR Code for facilitating the payment or fund transfer. Usually, an extra layer of security is supplemented by additional complex encryptions and tokenisation and/or other authentication techniques such as fingerprint face-recognition etc.

## III. MULTIFACETED SOCIAL MEDIA PAYMENT PLATFORMS

There is an emerging trend towards mobile payments moving into various social media platforms enabling p2p fund transfer amongst social media users. Examples of such platform includes: WeChat Pay, Venmo, Facebook Messenger, Google Wallet, Twitter and iPayYou.

### A. Venmo

The underlying payment protocol of Venmo is pay-by-text service i.e. SMS payment – enabling users to pay each other by sending SMS. Payees do not need to have prior registration for receiving funds, however, once the payment notification SMS is received they need to register to retrieve the fund transferred.

Until the sender' identity is verified, the transaction is capped at 299 USD per week. Once verification is done, this limit is lifted to 2999 USD per week with a maximum limit of single transaction of 2000 USD.

In fact, Venmo is a venture of Paypal. However, the Venmo app highly differs from Paypal, especially with regards to its embedded social aspect.

### B. Google Pay Send (Google Wallet)

Aligned with multifaceted online services, Google Pay Send, formerly known as Google Wallet is another popular service by Google. However, this service has now been merged with Google Pay. Funds can be transferred by using either standalone Google Wallet apps or any other integrated Google services such as Gmail. Integration with other Google services is a unique feature of Google Pay Send, compared with other available digital wallets and/or social media payment platforms. In fact, users who have a Gmail email account have already essentially signed up for Google Pay Send.

Google Pay allows users to link two different bank accounts with their Google Pay account. Google Pay holds the money within its balance for any particular user. Users can either keep using the money from the balance or withdraw to any of the two linked bank accounts.

## C. Twitter (Tweet Purchase and iPayYou)

Twitter's Tweet Purchase option enables businesses to advertise and sell their products and services directly from their tweets. Tweet Purchase is further enhanced by its matching platform facilities – once seller tweets regarding their services and products, the underlying algorithms predict cohort of possible buyers and feeds the tweets to them. If there is a match, the buyer can simply hit the "buy" button to avail the desired tweeted products or services.

iPayYou, also known as Pay by Twitter, facilitates bitcoin transactions. Once a Bitcoin payment transaction via iPayYou is completed, the recipient is notified through a tweet – containing a iPayYou website's link associated with the payment. If the receiver does not have an iPayYou account, the user will first have to register before the transferred Bitcoin can be used for any purposes – such as transfer the Bitcoin to someone else, purchase anything using the Bitcoin, convert the Bitcoin into US Dollars or even keep the Bitcoin for future use.

## D. WeChat Pay

WeChat – a multipurpose social media app, mainly extremely popular in Greater China region – is a venture of Tencent [3]. In addition to many other services, WeChat Pay is widely used in this region. WeChat pay utilises and offers a rich combination of various mobile payment protocols such as NFC, QR Code etc. While, users who link their debit cards to their WeChat accounts can use WeChat Pay for p2p fund transfers, purchasing goods and services, receiving funds etc., users who link their credit cards can only make payments.

WeChat HK, especially designed for the residents of Hong Kong (HK) with facilities to link local bank or credit cards, has now more than 40 business vendor partners in Hong Kong such as all the food chains under the Maxim Group and retail partners including G2000 apparel and Sasa Cosmetics. In fact, the popularity of WeChat Pay as well as AliPay is rapidly increasing in Hong Kong in recent years [3], as indicated by a survey conducted by GroupM [4], refer to figure 2. However, the small sample size of the survey needs to be considered before any generalisation of the findings.

## E. Facebook Messenger

In fact, payment facilities by Facebook have started long before the declaration of developing its cryptocurrency "libra". Facebook users first need to integrate their plastic card details or paypal account with their respective Facebook account and then make payments to other facebook users via Facebook Messenger. However, this facility is now limited to few countries such as USA (in US Dollars), France (in Euro) and United Kingdom (in Pound Sterling). Facebook Messenger enables the users to make payment to any other facebook users or request payment from someone they owe money or expect transfer of funds. It also facilitates to split any group payment amongst the participants such as having dinners with friends.

Before making the first payment, users require to provide Facebook with funding bank account or credit card or Facebook gift card details. The funds can be transferred by simply hitting the "$" sign, followed by the amount and then

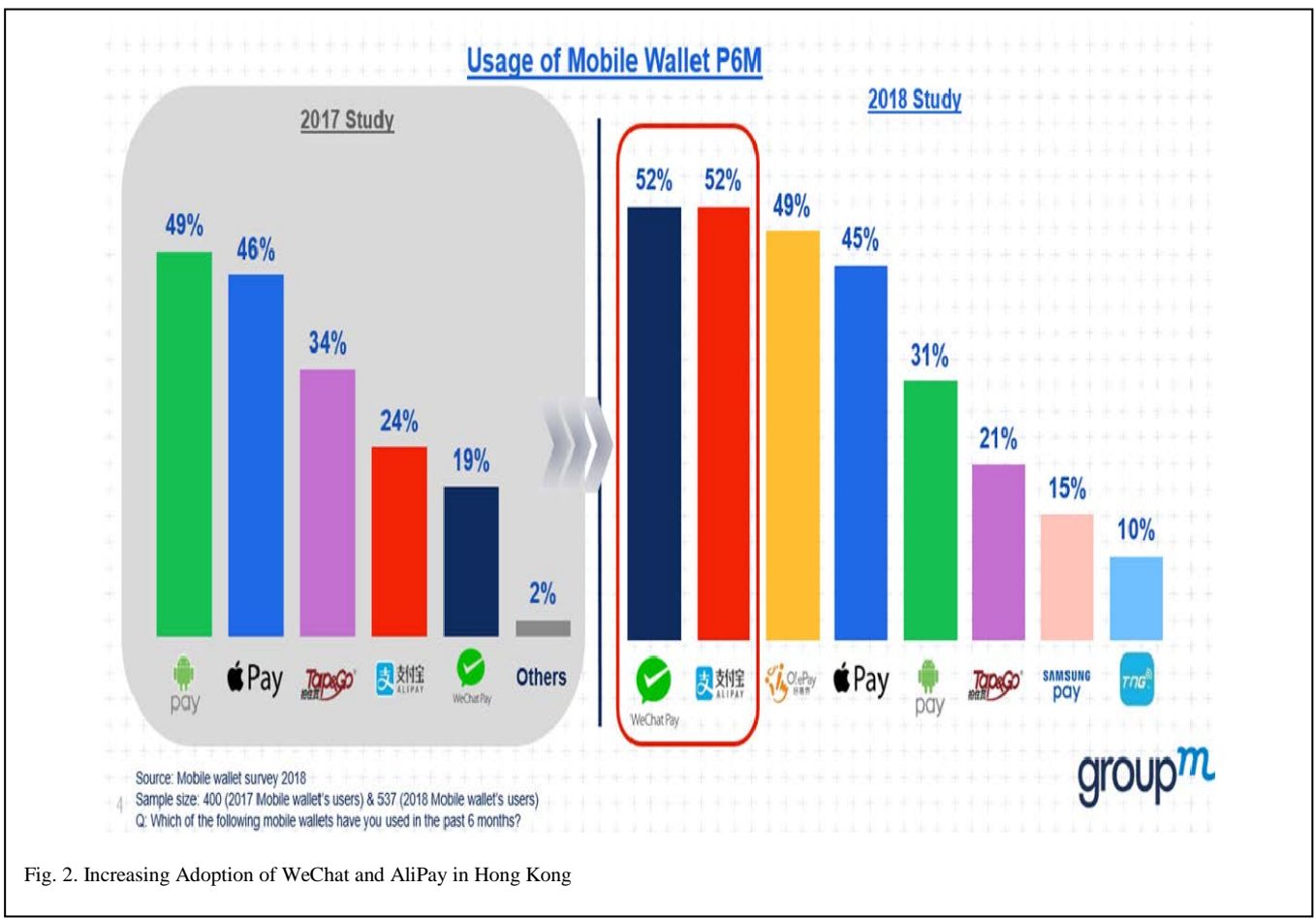

Fig. 2. Increasing Adoption of WeChat and AliPay in Hong Kong

pressing "Pay" sign. For additional security, a Personal Identification Number (PIN) can also be set up. The received funds are usually transferred to the associated checking account linked the debit card. While Facebook transfers the fund immediately without holding it, it may take some time before it is cleared by the recipient bank.

In fact, Facebook has recently (2019) declared implementing blockchain based cryptocurrency "Libra", governed by an independent Libra Association, using Byzantine Fault Tolerant Consensus. Libra will be backed by a reserve of assets in order to provide it with intrinsic value [5]. Either any crypto-exchange or Facebook's own digital wallet "Calibra" will be required for using Libra.

### F. Apple Pay
In Europe, Middle-East, United-States, and Asia, there is now a competitor developed by Apple called Apple Pay[2] . Users of Apple Pay can pay in shops, on mobile applications, and on Safari. There is the possibility to use Apple Pay internationally. However, in Asia, Apple Pay has many competitors like Ali Pay, WeChat, etc. The difference between Apple Pay and the competitors, is that Apple Pay is only supported by Apple devices. Apple Pay is used to withdraw cash directly from the bank ATM machine with an NFC device. The security is higher because of the fingerprints recognition for the Apple Pay than the competitors. Ali Pay and WeChat are leading in Asia, whereas, Apple Pay is more present internationally[3].

### IV. USERS' BEHAVIOUR AND MOTIVATIONS AND FOR ADOPTION

While social media payment platforms are different services than mobile payment platforms, there is a very deep interrelation, especially in terms of adoption and diffusion trends. Therefore, the scope of our literature survey also covers mobile payment aspects as an input for the prediction of social media payment platform's future adoption trends. In fact, while there were many surveys and research conducted in the domain of mobile payment platforms, social media aspects had little attention. Our current research project aims to fill in this gap. Furthermore, most of the research conducted on the adoption trends were mainly based on the Technology Acceptance Model (TAM) [6] or its extended version. In fact, Davies et al. [6] in 1989 introduced TAM adapting from the general Theory of Reasoned Action (TRA) model put forward by Fishbein and Ajzen in 1975 [7], to customise the TRA theory for studying adoption and diffusion trends of information systems. However, "Diffusion of Innovations" by Rogers [8-9] is another approach of measuring the adoption of technology diffusion. Later (in 1971), Rogers himself revamped "Diffusion of Innovations" theory into a more discretely selective version "Communication of Innovations; A Cross-Cultural Approach" [10] to include the effect of culture and societal norms in adoption and diffusion of any technological product and/or service. Therefore, we aim to rather use the later approach to investigate in a more meticulous approach considering the cultural [11-13] aspects too.

Mobile (or any other digital) payment platforms is gradually replacing traditional channel of payments. One of its great advantage is that moving of physical tokens or objects such as cash money, cheques, banker's draft or any other forms of assets is not required anymore. As a result, it not only saves time but also reduces the cost of transmitting these objects physically, securing, monitoring and other associated costs and hassles. Therefore, these new innovative channels are increasingly being adopted by the users.

These payment networks additionally provides some other benefits and motivations too, such as  cash back rewards, rewards for introducing new members, airline miles, redeemable royalty points and points toward hotel stays, to attract new customers as well as to increase participation [14]. Expanding customers' protection for any risks associated with non-delivery of the purchased product/service [15] and legislations – from within the industry- limiting consumer subjections to losses occurred from fraudulent charges [16] has further uplifted customers confidence in using them. These contactless payments are fast because the mobile holder just waves his phone in front of a reader in a shop. Another motivation is the fact that users can switch between their bank accounts and even benefit from PFM services to have an overview of their transactions, spendings, and manage their accounts (Ali Pay and WeChat) [17].

There are approximately 2.46 billion social media users globally[18]. Because of its extreme popularity and low cost solution, social media is now being utilised as a marketing tool, especially by customer-to-customer (C2C) ventures [19]. Introduction of Mobile Web 2.0, an evolution of Web 2.0 for mobile devices, the scope of social media and thus social commerce has expanded and migrated to mobile platform [18, 20].

Sajid and Haddara [21] conducted a study, using TAM approach, amongst Norwegian users which suggest that NFC based mobile payment is gradually being adopted. As per the findings of their research. Kalinica et al. [22] conducted a study with a sample size of 701 users, using structural equation modeling (SEM) for projection of P2P mobile payment platforms and then using neural network to rank the predictors obtained through SEM. While their research reveal "usefulness" to be the most significant factor for adoption, they advocate that perceived trust as well as social norms are also important.

Considering mobile payment platforms as a key driving factor of socioeconomic development, research of [23] advocates that technological advancements, socioeconomic circumstances and increasing use of mobile devices are playing important roles in the adoption of mobile payment platforms in some prominent markets.

Zmijewska [24] adopted a customer approach to investigate adoption of mobile payments platforms which identified six significant acceptance factors affecting user adoption:  perceived usefulness, ease of use i.e. usability, mobility, expressiveness, cost and perceived trust – usability being the most significant. One interesting finding of this research is preference of NFC over Bluetooth by the mobile payment users, as unlike Bluetooth, NFC does not require pairing or any other special setup. In fact, there are several

---

other research highlighted the advantages of NFC based mobile payment such as ubiquitousness [25-27], minimal user learning curve and anonymity [25], queue avoidance [26], convenience [28-29], enhanced security [30] and enriched user management for spending and finances [31]. However, lack of coherent ecosystems and standardisation are considered two major barriers for adoption of NFC [25-26, 32-34]. Unauthorised use of the NFC enabled device and security concerns are also to be considered as barriers to the adoption of mobile payment [32, 35].

Another study [36] used extended TAM with some additional factors, such as perceived compatibility, perceived trust and perceived risk, to further investigate the aspects and mechanisms affecting the adoption of mobile payment platforms. In this study [36], perceived risk was further divided into two: perceived information risk and perceived financial risk. The model was then tested by conducting a user survey of a sample size of 295. Their results identified perceived trust and perceived ease of use to be the two most significant factors affecting users' motivation for adopting mobile payment platforms. Another similar study [37] was conducted among Chinese users of Alipay – perceived ese of use and perceived usefulness were identified as having most significant contributions towards adoption of mobile payment systems.

Another study [38] was conducted using traditional TAM to identify the adoption factors of mobile payment platforms particularly in social networks i.e. social media payment platforms using mobile payment. In this study, users' age, gender and level of experience we analysed to establish the decisive factors amongst social network users. The study urged the need for implementing new business models adopting the new technological advancements.

A different approach was adopted by a study jointly conducted by Mastercard and PRIME Research [39] in 2014 which analyses 13 million social media posts and comments across 56 markets and 26 languages in North and South America, Europe, Africa, Asia and the Pacific Rim to assess people's use of products, adoption willingness and sentiment toward existing options. Utilising their proprietary social media analytics methods and technology, sentiment analysis reveals that the lion share of posts were motivated by news-story sharing. The study found improved sentiments towards adoption these new innovations within the payment industry.

Based on this literature review, this can be concluded that while there are many factors affecting the adoption of social media and mobile payment platforms, however, overall it demonstrates a positive sentiment in terms of adoption and acceptance. However, risk factors associated with monetary transactions in both fiat and digital forms are very high. Therefore, a carefully designed policy, based on a multidisciplinary approach combining technological, legal as well as regulatory aspects is required. Many other aspects such as social norms, socio-economical circumstances and cross-border cooperation are also to be considered.

## V. CONCLUDING DISCUSSION

This paper summarises the multifaceted features of social media and mobile payment solutions for better understanding of their rate of diffusion. The paper then analyses users' behaviour and forecast the future adoption trends of such platforms amongst various diverged users. The research also investigates privacy, security, regulatory as well as legal and ethical issues. Future research directions include conducting user surveys in various geographical areas as well as survey of legal and regulatory frameworks in different jurisdictions.

## ACKNOWLEDGMENT (*Heading 5*)

This research is supported by the Centre for Financial Regulation and Economic Development (CFRED), The Chinese University of Hong Kong (CUHK), Hong Kong SAR.